\documentclass[%
 reprint,
superscriptaddress,
 amsmath,amssymb,
 aps,
prl
]{revtex4-2}

\usepackage{amsthm}
\usepackage{mathrsfs}
\usepackage{systeme, mathtools}

\usepackage{xcolor}
\usepackage{graphicx}
\usepackage{dcolumn}
\usepackage{bm}

\newtheorem{myconj}{Conjecture}

\begin{document}

\title{Is the diagonal case a general picture for Loop Quantum Cosmology?}

\author{Matteo Bruno}
 \email{matteo.bruno@uniroma1.it}
 \affiliation{Physics Department, Sapienza University of Rome, P.le A. Moro 5, 00185 Roma, Italy}
\author{Giovanni Montani}
 \email{giovanni.montani@enea.it}
 \affiliation{ENEA, C.R. Frascati (Rome), Italy Via E.\ Fermi 45, 00044 Frascati (Roma), Italy} 
 \affiliation{Physics Department, Sapienza University of Rome, P.le A. Moro 5, 00185 Roma, Italy}

\begin{abstract}
    The correct implementation of the Loop Quantum Gravity to the early homogeneous Universe has been the subject of a long debate in the literature because the $SU(2)$ symmetry cannot be properly retained. The role of this symmetry is expressed by the Gauss constraint. Here, a non-vanishing Gauss constraint is found. However, we show that using suitable variables, it can be recast into three Abelian constraints, justifying the absence of such a symmetry in Loop Quantum Cosmology.
\end{abstract}

\maketitle

\section{Introduction}
The most promising proposal to quantize the gravitational field is, till now, the so-called Loop Quantum Gravity \cite{thiemann_2007,rovelli_2004,cianfrani2014canonical}. This claim is based on the idea that such a proposal, starting from a classical formulation of General Relativity, which is (on shell) equivalent to the Einstein-Hilbert formulation \cite{ashtekar86,ashtekar87, Holst1996, Barbero1995, Immirzi_1997}, arrives, via the introduction of the $SU(2)$ symmetry, to describe geometrical operators, like areas and volumes of space, as associated to discrete spectrum \cite{rovelli1995discreteness}. As a consequence, the implementation of Loop Quantum Gravity to the cosmological setting led to a Big-Bounce for the primordial Universe \cite{ashtekar2006quantumI,ashtekar2006quantumII,bojowald2002isotropic}, due to an anomaly of the classical limit.\\

The reliability of the so-called Loop Quantum Cosmology has been debated over the years \cite{CianfraniMontani2012gauge, CianfraniMontani2012critical,bojowald2020}, because the symmetry restriction induced by the homogeneity constraint prevents the preservation of the $SU(2)$ symmetry in the classical and quantum formulation. Actually, the its-self implementation of the dynamics for homogeneous models is a step forward in the general formulation, for which a reliable implementation of the regularized scalar constraint \cite{Thiemann_1998_I,Thiemann_1998_II} is not viable \cite{Nicolai_2005}.\\

An interesting attempt to restore also in cosmology a gauge $SU(2)$ symmetry, together with the associated Gauss constraint, has been formulated in Refs.~\cite{bojowald2000loopI,bojowald2000loopII,bojowald2000loopIII,bojowald2013mathematical}. There, a kinematical Hilbert space has been constructed by emulating the basic formulation in Loop Quantum Gravity. The idea is that the homogeneity of the space still allows for a local time-dependent Lorentz rotation of the triad vectors, so restoring a non-identically vanishing Gauss constraint as for the original formulation of the Ashtekar School \cite{ashtekar2011loop}.\\

The present analysis starts from the same theoretical set-up of a local time-dependent gauge transformation of the triad, but, investigating in detail the relation of the Ashtekar-Barbero-Immirzi connection and conjugate momentum to the standard ADM-Hamiltonian variables, it arrives at a rather different conclusion: the resulting picture is closer to the formulation of the Ashtekar School than to real "spin-network'' construction \cite{thiemann_2007}. When we express the $SU(2)$ gauge connection in terms of the metric variables (three scale factors, three Euler angles and, eventually three gauge angles), a local expansion of the involved functions outlines a linear dependence of the Gauss constraint from the three momenta variables, associated to the gauge angles (i.e. those responsible for the local Lorentz rotation). This result suggests pursuing, \emph{ab initio} a Holst formulation \cite{Holst1996}, by expressing the $SU(2)$ connection in terms of the 
metric variables. This calculus strategy provides the net and relevant issue of a linear relationship between the three Gauss constraint components and the three null momenta of the gauge angles: the Gauss constraint validity is ensured by the simultaneous vanishing behaviour of the three momenta and vice versa.\\

Particularly, we demonstrate that the Gauss constraints can be suitably restated into three Abelian constraints, simply stating the gauge nature of the three angles which rotates the dreibein. The explicit expression of the matrix linking the two sets of constraints is provided here.\\

Finally, the most important consequence of the present study is that the physical kinematical states of the theory cannot depend on the three gauge angles (simply because in a canonical formulation they are annihilated by the three null momenta) so that the quantization of the model reduces to the analysis provided in Ref.~\cite{io_1} on the non-diagonal Bianchi models. In other words, the present study allows validation of the original idea that the space of the almost-periodic functions is the suitable approach to implement a canonical Loop Quantum Gravity in cosmology. Even if we start with all the nine non-zero triad components, three of them are actually gauge angles, leading to a Gauss constraint that is reducible to three Abelian vanishing momenta. The quantization coincides with that one of a non-diagonal Bianchi Universe, which in Ref.~\cite{io_1} was associated with a diagonal representation of the fluxes, in agreement with the analysis in Refs.~\cite{ashtekar2003mathematical,ashtekar2009loop,ashtekar2007}.

\section{Rotations as gauge transformations}
We recall the classical description of Ashtekar variables in a homogeneous Universe. In a homogeneous model, the space-time is a manifold $\mathcal{M}\cong\mathbb{R}\times\Sigma$, where $\Sigma$ is a three-dimensional Riemannian homogeneous space. We require that the isometry group $S$ of $\Sigma$ acts transitively and freely \cite{bojowald2013mathematical}.\\
On $\Sigma$ exists a basis of left-invariant one-forms $\omega^I$ (i.e. $F^*\omega^I=\omega^I$, $\forall F\in S$) such that
\begin{equation}
    \label{MCeq}
    d\omega^I+\frac{1}{2}f^I_{JK}\omega^J\wedge\omega^K=0.
\end{equation}
The dual vector fields $\xi_I$ (defined by $\omega^I(\xi_J)=\delta^I_J$) are the generators of the Lie algebra $\mathfrak{s}$ of $S$
\begin{equation}
    \label{algebra}
    [\xi_I,\xi_J]=f^K_{IJ}\xi_K,
\end{equation}
thus, $f^K_{IJ}$ are the structure constants.\\
The induced Riemannian metric $h$ on $\Sigma$ is left-invariant due to the homogeneous hypothesis, hence, it can be written in terms of $\omega^I$
\begin{equation}
    h=\eta_{IJ}\omega^I\otimes\omega^J,
\end{equation}
where $\eta_{IJ}$ is a symmetric matrix constant on $\Sigma$.\\
A homogeneous connection $A$ on $\Sigma$ is determined by a linear map $\phi:\mathfrak{s}\to\mathfrak{su}(2)$ and it is written as $A=\phi\circ \theta_{MC}$, where $\theta_{MC}=\xi_I\otimes\omega^I$ is the Maurer-Cartan form \cite{bojowald2000loopI}.\\

Using a coordinate system $(t,x^i)$ adapted to the space-time decomposition, the components of the left-invariant one-forms $\omega^I_i$ and the dual vector fields $\xi_I^i$ depend only on $x^i$, while the other quantities that are constant on $\Sigma$ are functions on $t$. Thus, the Ashtekar variables read \cite{ashtekar2003mathematical}:
\begin{equation}
     A^a_i(t,x)=\phi^a_I(t)\omega^I_i(x),\ E^i_a(t,x)=|\mathrm{det}(\omega^J_j(x))|p^I_a(t)\xi^i_I(x).
\end{equation}
We can also characterize the space-time metric $g$ via its component: $g_{00}=-N^2+N^iN^jh_{ij},\ g_{0i}=N^jh_{ij},\ g_{ij}=h_{ij}$, where $N$ and $N^i$ are the lapse function and the shift vector, respectively, and $h$ is the induced Riemannian metric $h_{ij}=\eta_{IJ}(t)\omega^I_i(x)\omega^J_j(x)$. In a homogeneous model, the lapse function is a function of time only $N=N(t)$, while the shift vector can be factorized as $N^i=N^I(t)\xi^i_I(x)$.\\

Now, we are interested in the gauge freedom of the Ashtekar variables. The gauge transformation for the densitized triads is known $p^I_a\mapsto p^I_b O^b_a$, with $O\in SO(3)$ \cite{bojowald2013mathematical}.\\
Due to the homogeneity hypothesis, $p^I_a$ only depends on time and this property must hold also after the gauge transformation. Hence, although $O$ can be arbitrary and does not contribute in any physical sense, it must depend on time only too.\\
Moreover, the gauge transformation can be seen as a rotation of the dreibein $e^i_a\mapsto O^b_ae^i_b$. This interpretation allows us to find the associated gauge transformation of the connection variables $\phi^a_I$.\\
Consider the usual expression of the Ashtekar connection $A^a_i=\Gamma^a_i+\gamma K^a_i$ where $\gamma$ is the Barbero-Immirzi parameter. We can treat the two terms separately. The second term $K^a_i=K_{ij}e^{aj}$ contains the external curvature $K_{ij}$ which is a geometrical quantity and is not affected by gauge transformations, while $e^{ja}$ is a dreibein vector, so it rotates under a gauge transformation. It is easy to check that the rotation matrix is the inverse of the transformation matrix that acts on $e^i_a$ because $\delta^i_j=e^i_ae^a_j$ must be invariant. Then, $ K_{ij}e^{ja}\ \mapsto\ K_{ij}(O^{-1})^a_b e^{jb}$.\\
Moreover, also the spin part transforms as $\Gamma^a_i\mapsto(O^{-1})^a_b\Gamma^b_i$. Thus, under a gauge transformation, a matrix rotation appears:
\begin{equation}
     A^a_i=\phi^a_I\omega^I_i\mapsto\ (O^{-1})^a_bA^b_i=(O^{-1})^a_b\phi^b_I\omega^I_i.
\end{equation}
Therefore, on the phase space $(\phi^a_I,p^J_b)$ the gauge transformation acts as
\begin{equation}
    p^I_a\mapsto p^I_b O^b_a\ ,\ \ \phi^a_I\mapsto  (O^t)^a_b\phi^b_I.
\end{equation}
We can check that such a transformation leaves the Gauss constraint weakly vanishes. In fact, the transformation of the Gauss constraint $G_a=\epsilon_{ab}^{\ \ c}\phi^b_I p^I_c$ reads
\begin{equation}
    \label{weakvan}
    G_a\mapsto\epsilon_{ab}^{\ \ c}(O^t)^b_d O^e_c\phi^d_I p^I_e=\epsilon_{bd}^{\ \ e}O^b_a\phi^d_I p^I_e=O^b_a G_b\approx 0
\end{equation}

Now, we look for a description in metric variables like the ones in Ref.~\cite{io_1}. The new phase space of the metric variables, composed of the three scale factors $a,b,c$ and the three Euler angles of the physics rotation $\theta,\psi,\varphi$, needs to include variables of the gauge freedom.\\
Since $O\in SO(3)$, it can be written in terms of Euler angles
\begin{equation}
\label{gfree}
    O=\exp(\alpha j_3)\exp(\beta j_2)\exp(\gamma j_3),
\end{equation}
where $j_i$ are the real matrix generators of $SO(3)$. Then, the three gauge variables are these three Euler angles $(\alpha,\beta,\gamma)$, they are seen as a chart on $SO(3)$, $\alpha,\gamma\in(0,2\pi),\ \beta\in(0,\pi)$. Hence, the new configuration coordinates are $\{a,b,c,\theta,\psi,\varphi,\alpha,\beta,\gamma\}$.\\

In order to construct a theory in which the Gauss constraint does not vanish identically and in which the role of the cosmological quantities is made explicit, the assumption of phase space with configuration variables $\{a,b,c,\theta,\psi,\varphi,\alpha,\beta,\gamma\}$ seems to be a reasonable solution. The idea is to impose a canonical transformation between the two phase spaces such that the conjugate momenta to the gauge variables are included in the expression of the Ashtekar variables. These momenta will play a role in the Gauss constraint and they can be removed from the theory to recover the expressions in Ref.~\cite{io_1}.

\section{Examination of the Bianchi I model}
For the non-diagonal Bianchi I model, we have a simple expression of Ashtekar variables in terms of metric variables which allows us to do some computations. We want to use the connection and fluxes expression in Eqs. (39) and (42) from Ref.~\cite{io_1} properly gauge rotated
\begin{align}
\label{conn-gauge}
 &\phi^a_I=\frac{\gamma}{2Na_{b}}\Lambda^J_b\,\dot{\eta}_{JI}(O^t)^a_b\,,\\
 \label{flux-gauge}
 &p^I_a=a_ba_c\Lambda^I_dO^d_a\ \ \ \ \textrm{with $\epsilon_{abc}=1$}\,,
\end{align} 
where $\Lambda$ is the physical rotation, and $a_1,a_2,a_3$ are the scale factors.\\
A direct computation shows that, despite the gauge freedom, the Gauss constraint identically vanishes (as well as in Ref.~\cite{bojowald2003homogeneous}). Thus, the gauge momenta play a fundamental role in a non-vanishing Gauss constraint description. Now, we want to analyze this aspect.

\subsection{A Lagrangian approach}
\label{sec:LagApp}
We want to investigate what happens to the Hamiltonian formulation starting from the Hols action \cite{Holst1996,cianfrani2014canonical}
\begin{equation}
    \mathcal{S}_H=\frac{c^3}{8\pi G \gamma}\int dt d^3x\bigg(E^i_a\dot A^a_i+\lambda^aG_a-N^i\mathcal{V}_i-\frac{N}{2}\gamma\mathcal{S}\bigg),
\end{equation}
and considering the connection and the dreibein with respect to the metric variables (\ref{conn-gauge},\ref{flux-gauge}). Here, $\lambda^a$ are Lagrange multipliers and
\begin{align*}
    &G_a=\epsilon_{ab}^{\ \ c}\phi^b_I p^I_c\ ,\ \ \mathcal{D}_I=G_b\phi^b_I,\\
    &\mathcal{S}=-\frac{1}{\gamma^2|\mathrm{det}(p^K_c)|}\Big(p^I_a\phi^a_I p^J_b\phi^b_J-p^I_a\phi^a_J p^J_b\phi^b_I\Big),
\end{align*}
are the Gauss, Diffeomorphism and scalar constraints, respectively.\\
We recall that $\phi^a_I$ is computed from the usual expression of the connection $A^a_i=\Gamma^a_i+\gamma K^a_i$, while $p^I_a$ has the geometrical meaning as the homogeneous part of the dreibein vectors. We want the Holst action to be explicit in terms of metric variables, the calculation of the single terms provides that the Gauss constraint vanishes $G_a=0$, and so the Diffeomorphism constraint $\mathcal{V}_i=0$, while the scalar constraint has the same expression as in Eq. (46) from Ref~\cite{io_1}.\\
As we expect, the gauge freedom does not appear in the Lagrangian that is invariant under gauge transformation. Hence, the momenta can be computed and the gauge momenta are null (i.e. $p_{\alpha}=0,\ p_{\beta}=0,\ p_{\gamma}=0$), while the others are the same presented in Ref.~\cite{io_1}. We can now perform the Legendre transformation with Lagrangian multipliers $\lambda_i$ to find the Hamiltonian
\begin{equation}
H=\frac{c^3}{8\pi G}\left(\lambda_1p_{\alpha}+\lambda_2 p_{\beta}+\lambda_3 p_{\gamma}-\frac{N}{2}\mathcal{S}\right).
\end{equation} 
Thus, the theory of the non-diagonal Bianchi I model in metric variables is a constrained Hamiltonian theory with phase space
$(a,b,c,\theta,\psi,\varphi,\alpha,\beta,\gamma,p_a,p_b,p_c,p_{\theta},p_{\psi},p_{\varphi},p_{\alpha},p_{\beta},p_{\gamma})$
with four constraints
\begin{equation}
    p_{\alpha}\approx0\ ,\ \ p_{\beta}\approx0\ ,\ \ p_{\gamma}\approx0\ ,\ \ \mathcal{S}\approx0
\end{equation}
and with a Hamiltonian which is a linear combination of such constraints.\\

Notice that the same theory written in terms of connection and dreibein $(\phi^a_I,p^J_b)$ has four constraints given by $G_a\approx0$ and $\mathcal{S}\approx0$.\\
The scalar constraint is the same in both formulations in the sense that it is possible to switch from one to the other using transformation (\ref{conn-gauge}). This property does not hold for the Gauss constraint, which also in gauge variables vanishes. However, it is replaced by the three constraints on the pure gauge momenta.\\

We interpret this result as follows: the Gauss constraint after the canonical transformation becomes the gauge momenta constraint. In such a way, the dependence on the gauge momenta we introduce in the Ashtekar variables vanishes on the constraints' hypersurface, recovering the usual description.  

\section{Equivalence between Gauss constraint and pure gauge momenta}
\label{sec:G-p}
In this Section, we want to find an explicit expression for the Gauss constraint. Previously, we showed that there exists a relation between the Gauss constraint and the momenta constraint, which we interpreted as
\begin{equation}
    \label{eqcon}
    G_a=0\ \ \ \ \ \iff\ \ \ \ \ p_g=0
\end{equation}
where $g\in\{\alpha,\beta,\gamma\}$.\\
Such a relation is satisfied if the Gauss constraint is linearly dependent on pure gauge momenta $p_g$ only. For simplicity, it will be our ansatz. Thus, we enunciate the following conjecture
\begin{myconj}
\label{conj:Gauss}
The Gauss constraint depends on the gauge momenta via a $3\times3$ matrix $L_{ag}$:
\begin{equation}
    \label{conj}
    G_a=L_{ag}p_g,
\end{equation}
with $a$ is $SU(2)$ internal index and $g\in\{\alpha,\beta,\gamma\}$.
\end{myconj}

Using this ansatz, we can explicitly compute the coefficients of the linear combination without using $M$, nor an explicit expression of $\phi^a_I$. Let $p^I_a$ as in Eq. (\ref{flux-gauge}) and the gauge momenta $p_g$ given by the transformation that satisfies the Lie condition (i.e. $p_gdq_g=\phi^a_I d p^I_a$).\\
With this assumption, the Gauss constraint reads
\begin{align*}
G_a&=\epsilon_{ab}^{\ \ c}\phi^b_I p^I_c=\phi^b_I (p^{\mathrm{ph}})^I_{d}\epsilon_{ab}^{\ \ c}O^d_c\\
&=L_{ag}p_g=L_{ag}\phi^b_I\frac{\partial p^I_b}{\partial q_g}=\phi^b_I (p^{\mathrm{ph}})^I_{d}L_{ag}\frac{\partial O^d_b}{\partial q_g}.
\end{align*}
Where $(p^{\mathrm{ph}})^I_{d}$ is the physical part of the dreibein (i.e. the one not gauge rotated).
From this, we can derive the following equation
\begin{equation}
\label{Leq}
\epsilon_{ab}^{\ \ c}=L_{ag}(O^t)^c_d\frac{\partial O^d_b}{\partial q_g}.
\end{equation}
This equation is enough to fully characterize the matrix $L_{ag}$, in fact, the RHS has the same skew-symmetric property as the Levi-Civita symbol. Therefore, we obtain nine linear independent equations. The equations' system has nine equations in nine variables and the associated determinant is $\sin^3\beta$, so, it is non-degenerate. Hence, it exists one and only one solution. The solution $L_{ag}$ can be found easily and it reads
\begin{equation}
L_{ag}=\begin{pmatrix}
-\csc\beta\cos\gamma & \sin\gamma & \cot\beta\cos\gamma\\
\csc\beta\sin\gamma & \cos\gamma & -\cot\beta\sin\gamma\\
0 & 0 & 1\\
\end{pmatrix}
\end{equation}
Finally, the Gauss constraint can be written explicitly in terms of gauge momenta
\begin{equation}
\label{gauss}
G_a=\left(\begin{array}{c}
-\csc\beta\cos\gamma p_{\alpha}+ \sin\gamma p_{\beta}+ \cot\beta\cos\gamma p_{\gamma}\\
\csc\beta\sin\gamma p_{\alpha}+ \cos\gamma p_{\beta} -\cot\beta\sin\gamma p_{\gamma}\\
p_{\gamma}
\end{array}\right)
\end{equation}

The conjecture enables us to find the explicit dependence of the Gauss constraint on the gauge momenta. The matrix of coefficients is invertible since its determinant is $\mathrm{det}(L_{ag})=-\csc\beta$, then the equivalence condition (\ref{eqcon}) holds.\\
This explains the vanishing Gauss constraint in our initial approach, and in general, in the similar approaches of Loop Quantum Cosmology. In fact, the connection $A^a_i=\Gamma^a_i+\gamma K^a_i$ is reduced and, when it is described in terms of metric variables, it results in a function defined on the constraints' hypersurface, as well as the dreibein, and so, the linear dependence (\ref{conj}) implies that the Gauss constraint computed from such a connection must vanish.

\section{Gauss constraint as the generator of gauge transformations}
It is well known that the Gauss constraint $G_a$ is the generator of the gauge transformations on the phase space $(\phi^a_I,p^J_b)$ \cite{rovelli_2004,thiemann_2007,bojowald2000loopI,io_1}. This feature should hold in the new variables. Therefore, we want to compute the canonical Poisson brackets with respect to $\{\alpha,\beta,\gamma,p_{\alpha},p_{\beta},p_{\gamma}\}$ of the Gauss constraint in (\ref{gauss}). We obtain
\begin{equation}
    \{G_a,G_b\}=-\epsilon_{abc}G_c.
\end{equation}
The sign is not relevant. We expect that this formulation comes out from a canonical transformation in which the connection and dreibein are switched in the phase space, and then a sign in the Poisson bracket appears.\\

Hence, the Gauss constraint respects the $\mathfrak{su}(2)$-Lie algebra and generates the $SU(2)$ gauge transformations. Furthermore, it is linear in gauge momenta, so the hypersurface defined by $G_a=0$ is also described by $p_{\alpha}=p_{\beta}=p_{\gamma}=0$. Thus, the Gauss constraint is equivalent to three constraints on the momenta. Consequently, the generators of the gauge transformation can be decomposed into three generators which commute each other 
\begin{equation}
    \{p_{\alpha},p_{\beta}\}=\{p_{\alpha},p_{\gamma}\}=\{p_{\beta},p_{\gamma}\}=0.
\end{equation}

This decomposition is particularly useful in the simplification of the implementation of the Gauss constraint in a quantum theory.

\subsection{Quantum Gauss constraint}
In Ref.~\cite{io_1} is shown that the quantization of the non-diagonal Bianchi I model can be done in diagonal fluxes and angles variables. It is reasonable that a similar quantization can be provided for the other non-diagonal models, given a loop quantization of homogeneous Universes. However, to complete the description in the loop framework, we need to include the gauge transformations and a non-vanishing Gauss constraint.\\
Supposing that we have a quantization like in the non-diagonal Bianchi I case, it is enough to add the gauge variables to the phase space of the diagonal fluxes and angles. These gauge variables will be the Euler angles of the gauge rotation and they will be quantized independently (as the physical angles \cite{io_1}) via the Schr\"odinger picture. Thus, the wave functions are $\Psi(p_1,p_2,p_3,\theta_1,\theta_2,\theta_3,\alpha,\beta,\gamma)$, where $p_1,p_2,p_3$ are the diagonal fluxes and $\theta_1,\theta_2,\theta_3$ are the physical angles.\\
Moreover, the Hamiltonian (such as the Lagrangian) is independent of the gauge variables, hence the wave function factorizes $\Psi=\phi(\alpha,\beta,\gamma)\Phi(p_1,p_2,p_3,\theta_1,\theta_2,\theta_3)$.\\
On these functions, the action on the Gauss constraint is essentially a first-order derivative, so the imposition of the weak constraint $\hat G_a\Psi=0$ is equivalent to
\begin{equation}
    -i\hbar\frac{\partial \Psi}{\partial \alpha}=0,\ -i\hbar\frac{\partial \Psi}{\partial \beta}=0,\ -i\hbar\frac{\partial \Psi}{\partial\gamma}=0.
\end{equation}
The solution of this set of equations is trivial: $\phi(\alpha,\beta,\gamma)=const$. Thus, the Gauss constraint in this Hilbert space imposes the independence of the wave function on the gauge angles. Therefore, the kinematical Hilbert space for the non-diagonal Bianchi I model presented in Ref.~\cite{io_1} remains the same also including the gauge transformations.

\section{Concluding remark}
The analysis above deepens the idea proposed in Ref \cite{bojowald2013mathematical}, that a non-vanishing Gauss constraint can be restored also in the minisuperspace of a Bianchi model, as soon as the most general for the Ashtekar-Barbero-Immirzi connection is considered.\\

Actually, we interpret this general formulation in terms of the ADM metric variables. The introduction of gauge variables is responsible for restoring the $SU(2)$ symmetry and ensuring that the corresponding connection has to verify a Gauss constraint. However, the main result we obtained is that the components of such a Gauss constraint are linearly dependent on the three momenta corresponding to the gauge angles. Thus, in terms of metric variables, the $SU(2)$ symmetry reduces to the vanishing behaviour of these three momenta, i.e. it is, \emph{de facto} reduced to a set of Abelian constraints.\\
We also clarified how the non-commutative character of the Gauss constraint components is restored via the transformation linking the two representations, associated with the $SU(2)$ generators.\\

This issue has a deep impact on the Dirac quantization of the model since the three momenta operators, associated with the gauge angles, must annihilate the state function, which is therefore independent of such angles. Hence, our quantization of the model is equivalent to a non-diagonal quantum Bianchi cosmology, as discussed in Ref.~\cite{io_1}, especially concerning the kinematical Hilbert space structure. Since in Ref.~\cite{io_1}, the quantum picture is associated with a diagonal set of flux variables, plus the three Euler angles expected to be canonically quantized, the present analysis allows us to claim that the quantization of the Bianchi I model, discussed in Ref.~\cite{ashtekar2009loop}, see also Refs.~\cite{CianfraniMontani2012critical} for a critical revision, is actually a rather general formulation, the only available in a minisuperspace dynamics. In other words, the scale factors associated in a Bianchi cosmology to independent space directions are the most relevant subjects of a Loop Quantum Cosmology quantization procedure and are characterized by an almost-periodic functions representation.\\

The reason why the minisuperspace $SU(2)$ symmetry can be decomposed on an Abelian symmetry of the phase space kinematics is reliably due to the fact that for the space Ashtekar-Barbero-Immirzi connection, a local Lorentz rotation depending on time only retains a global character. Thus, a genuine $SU(2)$-formulation in the sense of Loop Quantum Gravity is still forbidden.

\bibliographystyle{unsrt}
\bibliography{biblio.bib}

\begin{thebibliography}{10}

\bibitem{thiemann_2007}
T.~Thiemann.
\newblock {\em Modern Canonical Quantum General Relativity}.
\newblock Cambridge Monographs on Mathematical Physics. Cambridge University
  Press, 2007.

\bibitem{rovelli_2004}
C.~Rovelli.
\newblock {\em Quantum Gravity}.
\newblock Cambridge Monographs on Mathematical Physics. Cambridge University
  Press, 2004.

\bibitem{cianfrani2014canonical}
F.~Cianfrani, O.~M. Lecian, M.~Lulli, and G.~Montani.
\newblock {Canonical Quantum Gravity: Fundamentals and Recent Developments}.
\newblock 2014.

\bibitem{ashtekar86}
A.~Ashtekar.
\newblock {New Variables for Classical and Quantum Gravity}.
\newblock {\em Phys. Rev. Lett.}, 57:2244--2247, Nov 1986.

\bibitem{ashtekar87}
A.~Ashtekar.
\newblock New hamiltonian formulation of general relativity.
\newblock {\em Phys. Rev. D}, 36:1587--1602, Sep 1987.

\bibitem{Holst1996}
S.~Holst.
\newblock Barbero's hamiltonian derived from a generalized hilbert-palatini
  action.
\newblock {\em Phys. Rev. D}, 53:5966--5969, May 1996.

\bibitem{Barbero1995}
J.~F. Barbero~G.
\newblock Real ashtekar variables for lorentzian signature space-times.
\newblock {\em Phys. Rev. D}, 51:5507--5510, May 1995.

\bibitem{Immirzi_1997}
G.~Immirzi.
\newblock Real and complex connections for canonical gravity.
\newblock {\em Classical and Quantum Gravity}, 14(10):L177, oct 1997.

\bibitem{rovelli1995discreteness}
C.~Rovelli and L.~Smolin.
\newblock Discreteness of area and volume in quantum gravity.
\newblock {\em Nuclear Physics B}, 442(3):593--619, 1995.

\bibitem{ashtekar2006quantumI}
A.~Ashtekar, T.~Pawlowski, and P.~Singh.
\newblock Quantum nature of the big bang.
\newblock {\em Physical review letters}, 96(14):141301, 2006.

\bibitem{ashtekar2006quantumII}
A.~Ashtekar, T.~Pawlowski, and P.~Singh.
\newblock Quantum nature of the big bang: improved dynamics.
\newblock {\em Physical Review D}, 74(8):084003, 2006.

\bibitem{bojowald2002isotropic}
M.~Bojowald.
\newblock Isotropic loop quantum cosmology.
\newblock {\em Classical and Quantum Gravity}, 19(10):2717, apr 2002.

\bibitem{CianfraniMontani2012gauge}
F.~Cianfrani and G.~Montani.
\newblock Implications of the gauge-fixing loop quantum cosmology.
\newblock {\em Phys. Rev. D}, 85:024027, Jan 2012.

\bibitem{CianfraniMontani2012critical}
F.~Cianfrani and G.~Montani.
\newblock A critical analysis of the cosmological implementation of loop
  quantum gravity.
\newblock {\em Modern Physics Letters A}, 27(07):1250032, 2012.

\bibitem{bojowald2020}
M.~Bojowald.
\newblock Critical evaluation of common claims in loop quantum cosmology.
\newblock {\em Universe}, 6(3), 2020.

\bibitem{Thiemann_1998_I}
T.~Thiemann.
\newblock {Quantum spin dynamics (QSD)}.
\newblock {\em Classical and Quantum Gravity}, 15(4):839, apr 1998.

\bibitem{Thiemann_1998_II}
T.~Thiemann.
\newblock {Quantum spin dynamics (QSD): II. The kernel of the Wheeler - DeWitt
  constraint operator}.
\newblock {\em Classical and Quantum Gravity}, 15(4):875, apr 1998.

\bibitem{Nicolai_2005}
H.~Nicolai, K.~Peeters, and M.~Zamaklar.
\newblock Loop quantum gravity: an outside view.
\newblock {\em Classical and Quantum Gravity}, 22(19):R193, sep 2005.

\bibitem{bojowald2000loopI}
M.~Bojowald.
\newblock {Loop quantum cosmology: I. Kinematics}.
\newblock {\em Classical and Quantum Gravity}, 17(6):1489, 2000.

\bibitem{bojowald2000loopII}
M.~Bojowald.
\newblock {Loop quantum cosmology: {II}. Volume operators}.
\newblock {\em Classical and Quantum Gravity}, 17(6):1509, 2000.

\bibitem{bojowald2000loopIII}
M.~Bojowald.
\newblock {Loop quantum cosmology: {III}. Wheeler-DeWitt operators}.
\newblock {\em Classical and Quantum Gravity}, 18(6):1055, 2001.

\bibitem{bojowald2013mathematical}
M.~Bojowald.
\newblock Mathematical structure of loop quantum cosmology: Homogeneous models.
\newblock {\em SIGMA. Symmetry, Integrability and Geometry: Methods and
  Applications}, 9:082, 2013.

\bibitem{ashtekar2011loop}
A.~Ashtekar and P.~Singh.
\newblock Loop quantum cosmology: a status report.
\newblock {\em Classical and Quantum Gravity}, 28(21):213001, 2011.

\bibitem{io_1}
M.~Bruno and G.~Montani.
\newblock Loop quantum cosmology of nondiagonal bianchi models.
\newblock {\em Phys. Rev. D}, 107:126013, Jun 2023.

\bibitem{ashtekar2003mathematical}
A.~Ashtekar, M.~Bojowald, and J.~Lewandowski.
\newblock Mathematical structure of loop quantum cosmology.
\newblock {\em Advances in Theoretical and Mathematical Physics},
  7(2):233--268, 2003.

\bibitem{ashtekar2009loop}
A.~Ashtekar and E.~Wilson-Ewing.
\newblock Loop quantum cosmology of bianchi type {I} models.
\newblock {\em Physical Review D}, 79(8):083535, 2009.

\bibitem{ashtekar2007}
A.~Ashtekar, T.~Pawlowski, P.~Singh, and K.~Vandersloot.
\newblock {Loop quantum cosmology of $k=1$ FRW models}.
\newblock {\em Phys. Rev. D}, 75:024035, Jan 2007.

\bibitem{bojowald2003homogeneous}
M.~Bojowald.
\newblock Homogeneous loop quantum cosmology.
\newblock {\em Classical and Quantum Gravity}, 20(13):2595, 2003.

\end{thebibliography}

\end{document}